\documentclass[twocolumn,preprintnumbers,amsmath,amssymb]{revtex4}

\usepackage{graphicx}
\usepackage{dcolumn}
\usepackage{bm}
\usepackage{times,amsmath,amssymb,mathrsfs}

\begin{document}

\title{Theory of cavity-enhanced spontaneous parametric downconversion}

\author{Y. Jeronimo-Moreno$^a$, S. Rodriguez-Benavides$^b$ and A.B. U'Ren$^b$}
\affiliation{$^a$Departmento de \'{O}ptica, Centro de Investigaci\'{o}n Cient\'{i}fica y Educacion Superior de Ensenada (CICESE), Baja California, 22860, M\'{e}xico \\ $^b$Instituto de Ciencias Nucleares, Universidad Nacional Aut\'{o}noma de M\'{e}xico, apdo. postal 70-543, Distrito Federal, 04510, M\'{e}xico }

\begin{abstract} In this paper we study photon pairs generated in a nonlinear cavity, composed of a nonlinear crystal surrounded by mirrors, by the process of spontaneous parametric downconversion.  We analyze two different regimes: singly-resonant cavities where the signal and idler modes are resonant, and doubly-resonant cavities where the pump mode is also resonant.    We present analytic expressions for the joint spectral amplitude in these two cases, and study the reduction in emission bandwidth as the cavity coefficient finesse is increased.  We also study the enhancement of the source brightness resulting from the presence of the cavity.
\end{abstract}

\maketitle

\section{Introduction}
Photon-pair sources produced by the process of spontaneous parametric downconversion (SPDC) have been used in many experiments ranging from fundamental tests of quantum mechanics to practical implementations of quantum-enhanced technologies.  In past work it has been demonstrated that engineering the phasematching properties of nonlinear crystals, together with the appropriate selection of the pump characteristics, can result in the emission of photon pairs with a wide range of defining features.   Thus, particular phasematching regimes have been exploited for the generation of on the one hand factorable\cite{mosley08}, and on the other hand ultra-broadband photon-pairs\cite{odonnell07}. However, phasematching in nonlinear crystals cannot typically be engineered for the emission of narrowband photon-pairs, which represents a critical component for the efficient coupling of single photons to specific transitions of atomic systems.  Even a very long nonlinear medium (for example in a fiber-based, spontaneous four wave mixing source) which leads to a phasematching function with narrowband features in the space of generated frequencies, does not typically lead to narrowband photon-pair emission.     In this paper we present a theoretical analysis of the process of spontaneous parametric downconversion in a nonlinear cavity, formed by a nonlinear medium surrounded by mirrors.  This physical system represents a viable method for attaining narrow-band photon-pair generation.

Mirrors placed around an SPDC source, so that photon-pair amplitudes from different passes through the nonlinear crystal can interfere with each other, can lead to the suppression or to the enhancement of photon pair generation~\cite{herzog94}.  The presence of a high-finesse Fabry Perot cavity around the nonlinear crystal leads to interference between multiple time-displaced photon-pair probability amplitudes, which results in the suppression of photon pair emission at certain frequencies, and to the enhancement of photon pair emission at other frequencies.  With an appropriate source design, this approach may be exploited to engineer a photon-pair source to emit at particular narrowband spectral regions of interest.  Typical SPDC sources can have a bandwidth in the region of tens to hundreds of nm.  This corresponds to in the region of six orders of magnitude greater bandwidth for SPDC photon pairs (around hundreds of THz) than the typical bandwidth of an atomic transition (around tens of MHz).  Building on work by K\"{o}nig et al.~\cite{konig05},  Fedrizzi et al.~\cite{fedrizzi07}, Kuklewicz et al.~\cite{kuklewicz06} and Wang et al.~\cite{wang04}, recent papers by Neergaard-Nielsen~\cite{neergard-nielsen07} and Haase et al.~\cite{haase08} have experimentally demonstrated SPDC sources with an emission bandwidth in the tens of MHz, tailored for specific atomic transitions.

The  generation of narrowband photon pairs through cavity-enhanced SPDC represents the starting point and motivation for the present paper. Narrowband emission of course leads to a corresponding effect in the time domain.  Indeed, the correlation time, i.e. the width of the time of emission difference between the signal and idler modes, is correspondingly broadened as the photon pair emission bandwidth is decreased.   This effect has been exploited in order to circumvent the  experimental difficulty of resolving the photon-pair correlation time with available photodetectors~\cite{ou99,lu00,andrews01}. Another promising aspect of cavity-enhanced SPDC is that under appropriate conditions it can lead to an enhancement of the rate of emission~\cite{hariharan00}.

In this paper we study the process of spontaneous parametric downconversion (SPDC) in a nonlinear cavity with a pulsed pump.  In realistic examples where the bandwidth of the photon-pair source is sufficiently narrow to match that of an atomic transition, we utilize a small-bandwidth limit of the analysis developed here: a regime where the pump bandwidth corresponds to the linewidth of a continuous-wave laser.   A basic assumption throughout this work is that the process remains spontaneous, i.e. that despite possible multiple passes through the cavity of the pump pulses and generated photon pairs, at most a single photon pair is created by each pump pulse.   We will study two different regimes for the non-linear cavity: i) the singly-resonant case, by which we mean that both signal and idler (which may be frequency degenerate or  non-degenerate) are resonant, and ii) the doubly-resonant case, by which we mean that in addition to the signal and idler photons, the pump is also resonant.  We will also explore the effect of the presence of the cavity on the emitted photon-pair flux, and will present a specific source design tailored for coupling to a particular atomic transition.

\section{Singly-Resonant Cavity}
\label{Sec:SRC}

In this section, we concentrate on cavity-enhanced SPDC in the singly-resonant case; this implies that both mirrors are perfectly transmissive for the pump mode.  Thus, in this regime each pump pulse interacts with the crystal only once, and the two-photon e mission characteristics are subsequently modified through multiple reflections of the signal and idler photons at the two cavity mirrors.

\begin{figure}[ht]
\begin{center}
 \includegraphics[width=3.4in]{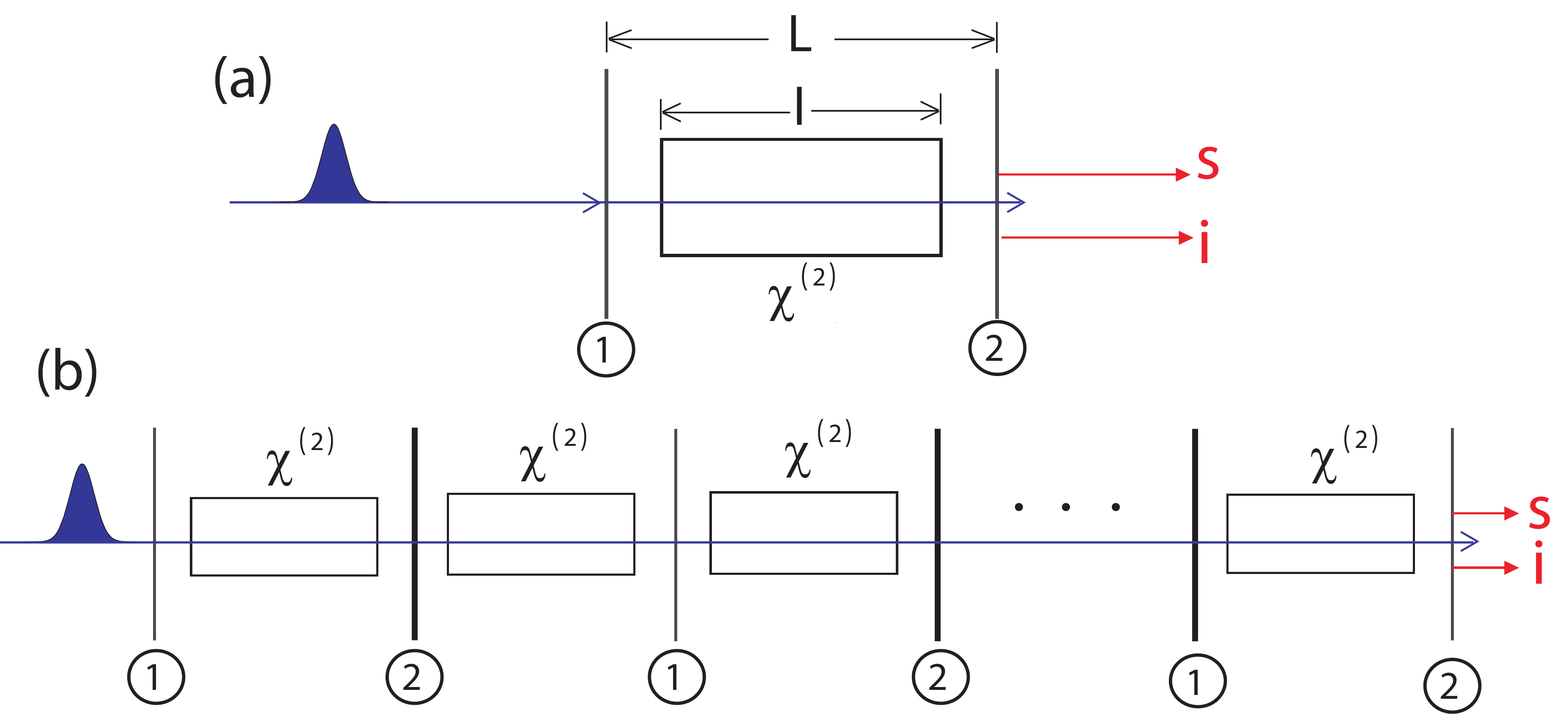}
 \end{center}
  \caption{(a) Schematic of a nonlinear cavity, showing the pump mode and the produced (signal and idler) modes, (b) Equivalent, `unfolded' cavity, used in
our calculations, where all waves propagate forward.}
  \label{Fg:cavSchem}
\end{figure}

Note that in general for each of the signal and idler modes, there will be two output modes i.e. backward-propagating, transmitted by the left-hand mirror and forward-propagating, transmitted by the right-hand mirror. In this paper we aim at capturing the essential physics of cavity-enhanced SPDC without the complexity of four output modes. Thus, we assume that the left-hand cavity mirror is perfectly reflective for the signal and idler modes, so that all emitted photons are forward-propagating, transmitted by the right-hand mirror.  Also, for simplicity, in this paper we concentrate on the frequency-degree of freedom.  A physical system to which this spectral-only description is directly applicable is  a cavity formed by a non-linear waveguide with a built-in mirror on either side (e.g. Bragg mirrors).  Fig.~\ref{Fg:cavSchem}(A) shows a schematic of the source under consideration; we label the left-hand and right-hand mirrors as $1$ and $2$ respectively and indicate the crystal length $\ell$ and cavity length $L$.  For $L>\ell$, we make the assumption that there are no reflections at the air-crystal and crystal-air interfaces.   Note that we can approach this situation using anti-reflection coatings.    In a realistic application it may be more practical to incorporate Bragg reflectors directly to the extrema of the nonlinear medium.

In the calculation presented below for generality we use different symbols for signal and idler quantities, in order to include in our description the frequency non-degenerate case.   We assume that mirror $1$ has an amplitude reflectivity $r_{1\mu}=\mbox{e}^{i\delta_{1\mu}}$ for each of the signal ($\mu=s$) and idler ($\mu=i$) modes, while mirror $2$ has amplitude reflectivity $r_{2\mu}=|r_{2\mu}|\mbox{e}^{i\delta_{2\mu}}$.  As has already been stated, we assume that the reflectivity for the pump pulses, in both mirrors, is zero.

We assume that the SPDC crystal fulfills type-I phasematching, with co-polarized signal and idler modes. In order to proceed with our analysis, we employ an equivalent physical system which consists of an unfolded cavity in which all waves propagate in the forward direction.   By making all waves forward-propagating, the cavity becomes a infinite sequence of crystals, each surrounded by two mirrors, as shown schematically in Fig.~\ref{Fg:cavSchem}(B).

At the second face of the first crystal in the unfolded cavity, the two-photon state is identical to that which would be observed without a cavity.   We may write down this state as follows

\begin{equation}
|\Psi \rangle = \int d \omega_s \int d \omega_i f(\omega_i,\omega_s) a_s^\dagger(\omega_s) a_i^\dagger(\omega_i) |\mbox{vac} \rangle \label{E:state}
\end{equation}

\noindent where $f(\omega_i,\omega_s)$ represents the joint spectral amplitude, and $a^\dagger_\mu(\omega_\mu)$ represents a creation operator at frequency $\omega_\mu$ for the intra-cavity mode (where $\mu=s,i$).  The joint spectral amplitude may be written in turn as $f(\omega_i,\omega_s)=\alpha(\omega_s+\omega_i) \phi(\omega_i,\omega_s)F_s(\omega_s) F_i(\omega_i)$ in terms of the pump envelope function $\alpha(\omega_s+\omega_i)$, the phasematching function $\phi(\omega_i,\omega_s)$, and spectral filter functions $F_\mu(\omega)$ (with $\mu=s,i$).  In this paper we model the pump envelope function as a Gaussian function $\alpha(\omega)=\exp[-(\omega-\omega_{p0})^2/\sigma^2]$ in terms of the pump central frequency $\omega_{p0}$ and the pump spectral width $\sigma$.

Immediately preceding the first mirror $2$ of the unfolded cavity, the single photon Fock state $|\omega \rangle_\mu$ appearing in Eq.~\ref{E:state} becomes

\begin{equation}
\label{ }
\left| \omega \right\rangle_\mu^{(1)-}=\mbox{e}^{i \gamma}a_{\mu}^\dag(\omega)\left|0 \right\rangle
\end{equation}

\noindent where $\gamma=\omega(L-l)/(2c)$ refers to the propagation phase over the segment of free space from the crystal's second face to mirror $2$.  The action of the first mirror $2$ is to split the intra-cavity mode $\hat{a}(\omega)$ into two modes, one of which corresponds to the same intra-cavity mode   $\hat{a}(\omega)$ and the second of which, to be referred to as $\hat{b}(\omega)$, refers to the transmitted, free-space mode.  Thus, the single photon Fock state immediately following the first mirror $2$ becomes

\begin{equation}
\label{E:JSASC}
\left| \omega \right\rangle_\mu^{(1)+}=\mbox{e}^{i \gamma}[r_{2\mu} a_{\mu}^\dag(\omega) +t_{2\mu} b_{\mu}^\dag(\omega)]\left|0 \right\rangle
\end{equation}

\noindent in terms of corresponding transmission $t_{\nu \mu}$ and reflection $r_{\nu\mu}$  amplitude coefficients for mirror $\nu$ (with $\nu=1,2$) and for mode $\mu$ (with $\mu=s,i$).  Carrying out an analysis for the propagation between subsequent mirrors $2$ and transmission/reflection at each mirror $2$, we arrive at the following expression for the state into which the original signal/idler single-photon Fock has evolved, immediately following the $n$th mirror $2$

\begin{align}
\label{E:npasos}
\left| \omega \right\rangle_\mu^{(n)}=&\mbox{e}^{i \gamma} \bigg[ r_{1\mu} ^{n-1}r_{2\mu} ^n \mbox{e}^{i2(n-1)\theta_\mu}a_{\mu}^\dag(\omega) + \\ \nonumber &
t_{2\mu} \sum_{j=0}^{n-1} r_{1\mu} ^{j} r_{2\mu} ^{j} \mbox{e}^{i(n-1-j)\Gamma_\mu}\mbox{e}^{i2j\theta_\mu}b_{\mu}^\dag(\omega)\bigg]\left|0 \right\rangle
\end{align}

\noindent where $\Gamma_\mu$ represents the phase accumulated by a photon $\mu$ in the extra-cavity mode during the period of time taken by an intra-cavity-mode photon to traverse the unfolded cavity from a mirror $2$ to the subsequent mirror $2$. In the specific case where $L=\ell$, this coefficient is given by $\Gamma_\mu=2 k^{\prime}(\omega_\mu)L$, where ${}^\prime$ denotes a frequency derivative. The quantity $\theta_\mu=\omega_\mu(L-\ell)/c + \ell n_\mu(\omega_\mu) \omega_\mu /c$, where the frequency dependence is implicit, refers to the phase accumulated during one pass of photon $\mu$ through the nonlinear crystal.

Carrying out the sum in Eq.~\ref{E:npasos} analytically, we can write the component of the quantum state for which both photons are in the extra-cavity mode after $n$ iterations of the cavity as

\begin{equation}
|\Psi^{(n)} \rangle = \int d \omega_s \int d \omega_i f^{(n)}_{\scriptsize{\textrm{SR}}}(\omega_i,\omega_s) b_s^\dagger(\omega_s) b_i^\dagger(\omega_i) |\mbox{vac} \rangle
\end{equation}

\noindent in terms of cavity-modified joint spectral amplitude function $f^{(n)}_{\scriptsize{\textrm{SR}}}(\omega_i,\omega_s)$

\begin{equation}
f^{(n)}_{\scriptsize{\textrm{SR}}}(\omega_i,\omega_s)=f(\omega_i,\omega_s)C^{(n)}_{\scriptsize{\textrm{SR}}}(\omega_i,\omega_s),
\end{equation}

\noindent in turn defined in terms of function $C^{(n)}_{\scriptsize{\textrm{SR}}}(\omega_i,\omega_s)$

\begin{align}
\label{ }
C^{(n)}_{\scriptsize{\textrm{SR}}}(\omega_i,\omega_s)&=A_s(\omega_s)A_i(\omega_i)
\end{align}
where
\begin{align}
\label{ }
A_\mu(\omega)=&t_ {2\mu} \mbox{e}^{i\left[ \gamma +(n -1)\Gamma_\mu\right]}\times \\ \nonumber & \frac{1-\left[   |r_{2\mu} |  \mbox{e}^{i(2\theta_\mu-\Gamma_\mu+\delta_{1\mu}+\delta_{2\mu})}\right]^n}{1-{  |r_{2\mu}| \mbox{e}^{i(2\theta_\mu-\Gamma_\mu+\delta_{1\mu}+\delta_{2\mu})}}}.
\end{align}

For the analysis which follows, we are interested in the quantum state corresponding to the limit where both photons are in the extra-cavity mode: that is, we consider a sufficient number of reflections (cavity iterations) so that the amplitudes which involve at least one intra-cavity photon are negligible.   Mathematically, this corresponds to the limit $n \rightarrow \infty$. In this limit, we may obtain an expression for the resulting joint spectral intensity.

\begin{align}
\label{E:fcavs}
&S_{\scriptsize{\textrm{SR}}}(\omega_i,\omega_s)=\lim_{n \rightarrow \infty}|f_{\scriptsize{\textrm{SR}}}^{(n)}(\omega_i,\omega_s)|^2 \nonumber \\
&=\mathscr{A}_s(\omega_s)\mathscr{A}_i(\omega_i) |f(\omega_i,\omega_s)|^2
\end{align}

\noindent written in terms of the Airy function $\mathscr{A}_\mu(\omega)$ (with $\mu=s,i$)

\begin{equation}
\label{E:Airy}
\mathscr{A}_\mu(\omega)=\frac{|t_{2\mu}|^2}{(1-|r_{2\mu} |)^2}\cdot\frac{1}{1+\mathscr{F}_\mu \sin^2(\Delta_\mu(\omega)/2)},
\end{equation}

\noindent defined in turn in terms of the coefficient of finesse $\mathscr{F}_\mu$, given by

\begin{equation}
\label{E:finesse}
\mathscr{F}_\mu=\frac{4|r_{2\mu}|}{(1-|r_{2\mu}|)^2}
\end{equation}

\noindent and the phase factor $\Delta_\mu(\omega)$, given by

\begin{equation}
\label{ }
\Delta_\mu(\omega)=2\theta_\mu+\delta_{1\mu}+{\delta_{2\mu}}-\Gamma_\mu.
\end{equation}

The Airy function $\mathscr{A}_\mu(\omega)$ (Eq.~\ref{E:Airy}) yields a sequence of equal-height peaks with width $\delta \omega$ and spectral separation $\Delta \omega$, also referred to as free spectral range.  Note that the reflectivity does not appear squared in Eq.~\ref{E:finesse} as in typical Fabry-Perot cavity treatments because in our case one of the two cavity mirrors is assumed to be perfectly reflective.  At each of the resulting peaks of function $\mathscr{A}_\mu(\omega)$, the cavity is resonant for photon $\mu$.  Note that if resonance is not achieved at the desired frequencies of operation, it is possible to adjust the reflection phases $\delta_{1 \mu}$ and  $\delta_{2 \mu}$ to ensure resonance.  Under the approximation that the index of refraction remains constant between the central peak (at $\omega_s=\omega_i=\omega_0$) and the subsequent (or preceding) peak, we obtain the following expressions for $\delta \omega$ and $\Delta \omega$

\begin{equation}
\label{Eq:deltaomega}
\delta\omega=\frac{2c}{l n(\omega_0)+(L-l)}\mathscr{F}^{-1/2}
\end{equation}

\noindent and

\begin{equation}
\label{Eq:Deltaw}
\Delta\omega=\frac{\pi c}{l n(\omega_0)+(L-l)},
\end{equation}

\noindent where $n(\omega_0)$ represents the index of refraction evaluated at the central SPDC frequency.

\begin{figure}[h]
 \includegraphics[width=3.4in]{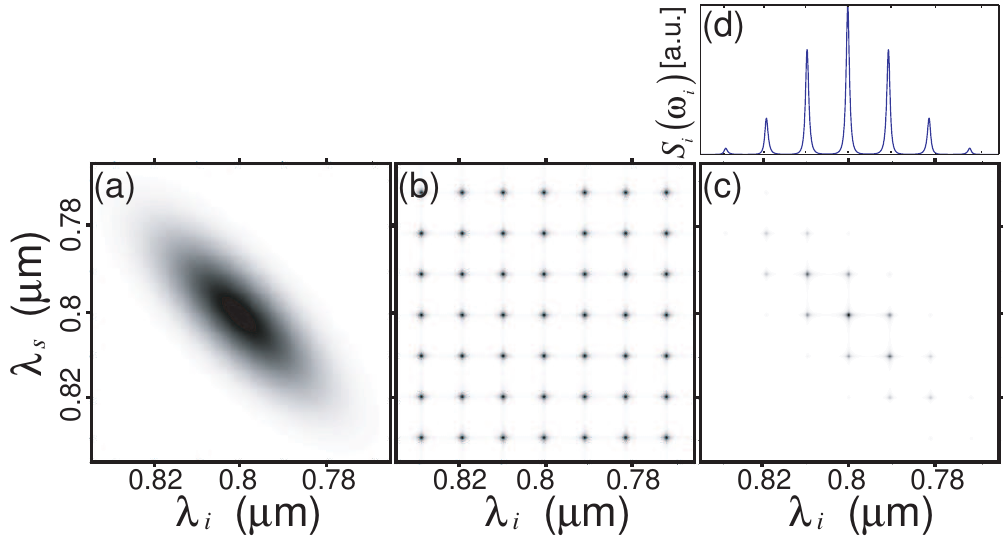}
\caption{(a) Joint spectral intensity for the equivalent source without a cavity, (b) Function $\mathscr{A}_s(\omega_s)\mathscr{A}_i(\omega_i)$, which describes the effect of the cavity, (c) Cavity-modified joint spectral intensity, and (d) Spectrum of signal-mode single photons.}
\label{Fg:JSISC}
\end{figure}

In order to illustrate the preceding discussion, we consider here a specific example.   We consider a nonlinear cavity formed by a beta-barium-borate (BBO) crystal of thickness $L=20\mu$m, within a cavity of the same length, i.e. with $L=\ell$. We assume that the pump is centered at $400$nm, with a full width at half maximum of $5$nm, while the SPDC photon pairs are centered at $800$nm.  For this illustration, we assume that while the reflectivity of mirror $1$ (at the SPDC frequency) is unity, the reflectivity of mirror $2$ is $|r_2|=0.73$.   Although a relatively low mirror 2 reflectivity value is assumed here for graphical clarity in the figure which follows, in a realistic cavity-based, narrow-band SPDC source design (see section~\ref{Sec:SpDesign}), a reflectivity value much closer to unity is likely to be required. Also note that for the short crystal length considered here, the bandwidth of the cavity-allowed modes $\delta \omega$ is much smaller than that of the phasematching function $\phi(\omega_i,\omega_s)$; in this illustration we have assumed the presence of a spectral filter, with a Gaussian transmissivity function and full width at half maximum $\Delta \lambda =30$nm, along the path of the SPDC photons following the non-linear cavity.  We have selected values for the reflection phases $\delta_{\nu \mu }$ (with $\nu=1,2$ and $\mu=s,i$) to ensure that the phase factor $\Delta_\mu(\omega)$ vanishes at the central signal and idler frequencies. Fig.~\ref{Fg:JSISC}(A) shows a plot of the joint spectral intensity function $|f(\omega_i,\omega_s)|^2$, i.e. which would be obtained without a cavity.  Fig.~\ref{Fg:JSISC}(B) shows a plot of $\mathscr{A}(\omega_s)\mathscr{A}(\omega_i)$; this function describes the effect of the interference which occurs due to multiple photon-pair reflections, on the resulting cavity-modified joint spectral intensity.   As can be observed, $\mathscr{A}(\omega_s)\mathscr{A}(\omega_i)$ corresponds to a square array of well-defined spectral cavity-allowed modes, with spectral width $\delta \omega$ and with separation between modes $\Delta \omega$.    Fig.~\ref{Fg:JSISC}(C) shows the cavity-modified joint spectral intensity $S_{\scriptsize{\textrm{SR}}}(\omega_i,\omega_s)$ (see Eq.~\ref{E:fcavs}).  This mode structure is also apparent in the marginal distribution $\int d \omega_s S_{\scriptsize{\textrm{SR}}}(\omega_i,\omega_s)$ [shown in Fig.~\ref{Fg:JSISC}(D)], which represents the spectrum of single photons in the signal mode.   As can be seen from the preceding two plots, the effect of the cavity on the two-photon state is to select those pairs of frequencies which correspond to cavity-allowed modes.

The main motivation for our work is to develop sources of nearly monochromatic photon pairs, where the signal and/or idler photons can be well-matched in central frequency and in bandwidth to specific atomic transitions.    As can be seen from Eq.~\ref{Eq:deltaomega}, the bandwidth of cavity-allowed modes scales with the cavity coefficient of finesse $\mathscr{F}$ as $\mathscr{F}^{-1/2}$.  Thus, a large coefficient of finesse, which corresponds to a high-quality cavity, leads to an important reduction in SPDC bandwidth.  We also note that in the case where the cavity and crystal have equal lengths ($L=\ell$), both the mode spectral width $\delta \omega$ and the spectral separation between modes $\Delta \omega$ scale with cavity length $L$ as $L^{-1}$.   In a practical implementation, it is desirable that a single cavity mode can be isolated, so that an idler detection event unambiguously heralds a single photon in the intended cavity mode. Shorter cavities imply a larger separation between modes $\Delta \omega$, which facilitates the isolation of a single cavity mode by transmitting the photon pairs through an appropriate narrow-band filter.   On the other hand, because flux scales with cavity length it is from that perspective desirable to make the cavity length as large as possible.  With this in mind, we propose a two-stage recipe for designing a nonlinear cavity source of photon pairs.  i) The minimum separation between cavity modes that permits the isolation of a single cavity mode sets the cavity length (with $L=\ell$), through Eq.~\ref{Eq:Deltaw}.  ii)  The required photon-pair bandwidth (e.g. determined by the bandwidth of the relevant atomic transition) determines the coefficient of finesse  $\mathscr{F}$ (which for frequency degenerate, type-I SPDC will be the same for both generated photons, i.e. $\mathscr{F} \equiv \mathscr{F}_s=\mathscr{F}_i$,  through Eq.~\ref{Eq:deltaomega}.

\begin{figure}[h!]
 \includegraphics[width=3.4in]{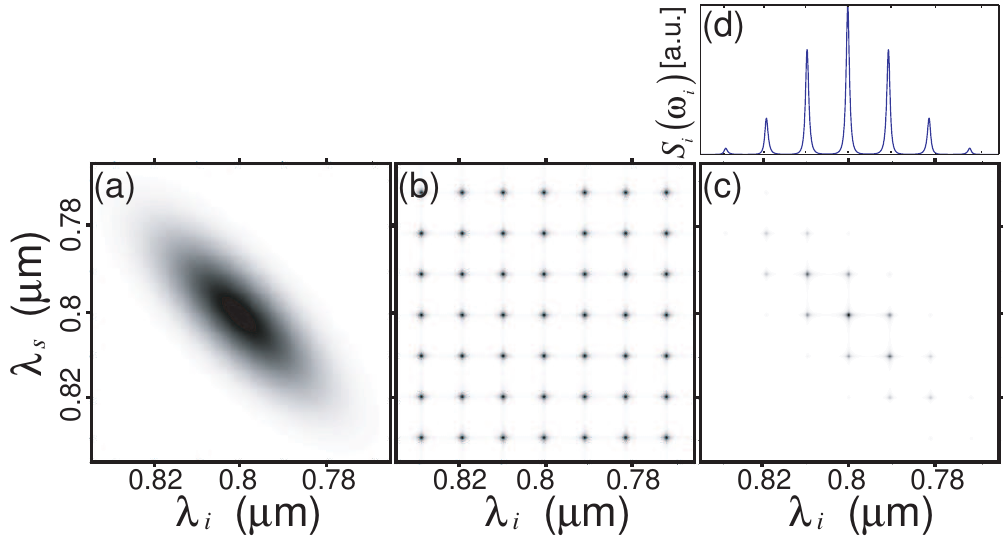}
\caption{(a) Distribution of time of emission differences between the signal and idler modes, showing well defined peaks where the peak separation corresponds to the cavity round-trip time. (b) Shown as a function of mirror $2$ reflectivity: i) spectral width of the cavity-allowed modes, and ii) photon-pair correlation time.}
\label{Fg:Corrtimedeltw}
\end{figure}

It is to be expected that in making the photon pairs increasingly narrow-band through a high-finesse nonlinear cavity, there will be a corresponding effect in the time domain.  In order to investigate this effect, let us consider the joint temporal amplitude $\tilde{f}_{\scriptsize{\textrm{SR}}}(t_{+},t_{-})$, defined as the Fourier transform of the joint spectral amplitude ${f}_{\scriptsize{\textrm{SR}}}(\omega_{+},\omega_{-})$, expressed in terms of variables $t_\pm$, which are the Fourier conjugate variables of $\omega_\pm \equiv \omega_s \pm \omega_i$.  From this, we obtain the joint temporal intensity $|\tilde{f}_{\scriptsize{\textrm{SR}}}(t_{+},t_{-})|^2$ and the distribution of emission difference times between the signal and idler modes, which corresponds to the marginal distribution $S_{-}(t_{-})=\int d t_{+} |\tilde{f}_{\scriptsize{\textrm{SR}}} (t_{+},t_{-}) |^2$.   The presence of the cavity implies that the signal and idler photons may be emitted at distinct times, corresponding to a different number of passes within the cavity.  This is illustrated in Fig.~\ref{Fg:Corrtimedeltw}(A), which shows a plot of $S_{-}(t_{-})$, calculated numerically, for the experimental parameters of Fig.~\ref{Fg:JSISC}.  This plot is composed of a series of peaks, where the peak separation corresponds to the cavity round trip time (at the SPDC frequency).   The integer number label corresponds to the the number of cavity round trip times by which the idler photon emission from the cavity precedes the signal photon emission from the cavity.  Note that, while the highest peak occurs at $t_{-}=t_s-t_i=0$, for a high-finesse cavity the heights of peaks will decay slowly, away from $t_{-}=0$.  We define the photon-pair correlation time $t_C$ as the width, expressed as the standard deviation, of the data set obtained from the peak maxima.

A higher coefficient of finesse (or higher mirror 2 reflectivity) will clearly result in a greater number of signal and idler photon passes through the cavity, translating into a more gradual reduction in the heights of peaks in the distribution of emission difference times.   This leads us to a discussion of the main cost of approaching monochromatic photon pair emission through a high-finesse nonlinear cavity: as the finesse is increased, the correlation time $t_C$ also increases.     In the monochromatic limit which corresponds to a perfect cavity with $\mathscr{F} \rightarrow \infty$, the photon pair character of the emitted modes is lost, i.e. the signal and idler modes are no longer temporally synchronized.    This behavior is clear in Fig.~\ref{Fg:Corrtimedeltw}(B) which shows: i) the spectral width (full width at half maximum) of the central cavity-allowed mode $\delta \omega$, and ii) the photon-pair correlation time $t_C$.   We also note that a very short correlation time (e.g. at the femtosecond level) may be difficult to resolve experimentally with current technology.  A nonlinear cavity may be used as an effective tool for the lengthening of this correlation time to a point where it may be experimentally resolved.

As has been discussed above, the need to isolate a single cavity mode leads to an upper limit for the nonlinear cavity length, which in turn limits the source brightness.    One way to transcend this limitation is by letting each pump pulse traverse the cavity multiple times, with each pass contributing to the source brightness.  Thus, in the next section we consider doubly-resonant cavities, i.e. which are resonant for both the SPDC and pump modes.

\section{Doubly-Resonant Cavity}
\label{Sec:CDR}

In this section we analyze the two-photon state produced by a cavity which is resonant for both, the SPDC and pump modes.  Later in this paper we will concentrate for simplicity on the case where the mirror $2$ has a perfect reflectivity for the pump (and mirror $1$ has a non-unity reflectivity) so that the pump pulses enter the nonlinear cavity from the left through mirror $1$, and the generated photons exit the cavity through mirror $2$ to the right.  However, we present here an analysis which is valid for arbitrary pump mode mirror $1$ and mirror $2$ reflectivities.  The pump-mode reflectivities are written as $r_{1p}=|r_{1p}|\exp(i \delta_{1p})$ and $r_{2p}=|r_{2p}|\exp(i \delta_{2p})$.

As in the previous section, it is helpful to consider an unfolded cavity, where the different sections of the unfolded cavity now correspond to each of the pump pulse passes through the cavity. Each pass of the pump pulse through the cavity leads to an amplitude $g_{\scriptsize{\textrm{DR}}}^{(j)}(\omega_i,\omega_s)$, so that the `total' joint amplitude resulting from all passes can be written as a sum over these individual amplitudes.    Amplitude $g_{\scriptsize{\textrm{DR}}}^{(j)}(\omega_i,\omega_s)$ incorporates the phase accumulated by the pump as it traverses the cavity $j$ times, as well as the corresponding reduced pump pulse amplitude for the $j$th pass (through a power of $r_{1p}r_{2p}$). For even $j$, it also includes the phase accumulated by the signal and idler photons as they traverse the cavity once so as to reach the next mirror $2$ in the unfolded cavity.   The amplitude $g_{\scriptsize{\textrm{DR}}}^{(j)}(\omega_i,\omega_s)$ incorporates function $f_{\scriptsize{\textrm{SR}}}(\omega_i,\omega_s)$ [defined as the limit $n \rightarrow \infty$ of $f_{\scriptsize{\textrm{SR}}}^{(n)}(\omega_i,\omega_s)$],  which describes the propagation of the signal and idler photons from the $j$th section of the unfolded cavity, onwards.

It is convenient to group the amplitudes $g_{\scriptsize{\textrm{DR}}}^{(j)}(\omega_i,\omega_s)$ as follows

\begin{eqnarray}
\label{ }
g_{\scriptsize{\textrm{DR}}}^{(1)}(\omega_i,\omega_s)&=&t_{1p}\mbox{e}^{i\gamma_p}f_{\scriptsize{\textrm{SR}}}(\omega_i,\omega_s)
\nonumber \\
g_{\scriptsize{\textrm{DR}}}^{(2+3)}(\omega_i,\omega_s)&=&t_{1p}r_{1p}r_{2p}\mbox{e}^{i\gamma_p}\mbox{e}^{i2\theta_p}Yf_{\scriptsize{\textrm{SR}}}(\omega_i,\omega_s)
\nonumber \\
g_{\scriptsize{\textrm{DR}}}^{(4+5)}(\omega_i,\omega_s)&=&t_{1p}r_{1p}^2r_{2p}^2\mbox{e}^{i\gamma_p}\mbox{e}^{i4\theta_p}Yf_{\scriptsize{\textrm{SR}}}(\omega_i,\omega_s)
\nonumber \\
g_{\scriptsize{\textrm{DR}}}^{(6+7)}(\omega_i,\omega_s)&=&t_{1p}r_{1p}^3r_{2p}^3\mbox{e}^{i\gamma_p}\mbox{e}^{i6\theta_p}
Yf_{\scriptsize{\textrm{SR}}}(\omega_i,\omega_s)
\end{eqnarray}

\noindent where higher-order terms follow the above pattern, and in terms of the following definitions

\begin{eqnarray}
Y &\equiv& 1+ r_{1p}^{-1} r_{1s}r_{1i} e^{i(\theta_{si}-\theta_p)}  \\
g_{\scriptsize{\textrm{DR}}}^{(j+[j+1])}(\omega_i,\omega_s) &\equiv& g_{\scriptsize{\textrm{DR}}}^{(j)}(\omega_i,\omega_s)+g_{\scriptsize{\textrm{DR}}}^{(j+1)}(\omega_i,\omega_s).
\end{eqnarray}

In the above expressions, $t_{1p}$ represents the pump-mode transmissivity of mirror $1$, $\gamma_p$ represents the phase accumulated by the pump pulse in traversing the empty space between mirror $1$ and the first face of the crystal, $\theta_p=\omega_p(L-l)/c+k_p(\omega_p) l$ is the phase accumulated by the pump in traversing the cavity once, and $\theta_{si}\equiv\theta_s+\theta_i$.  We then arrive at the following expression for the joint spectral amplitude where we have taken into account the first $1+2n$ passes of the pump through the cavity.

\begin{align}
&f_{\scriptsize{\textrm{DR}}}^{(1+2n)}(\omega_i,\omega_s)=\sum\limits_{j=1}^{1+2n} g^{(j)} _\textrm{DR}(\omega_i,\omega_s) \nonumber \\
&= t_{1p}\mbox{e}^{i\gamma_p}\bigg[1
+\Big(1+r_{1p}^{-1}r_{1s}r_{1i}\mbox{e}^{i(\theta_{si}-\theta_p)}\Big) \nonumber \\ &\times r_{1p}r_{2p}\mbox{e}^{i2\theta_p}\frac{1-(r_{1p}r_{2p}\mbox{e}^{i2\theta_p})^n}{1-r_{1p}r_{2p}\mbox{e}^{i2\theta_p}}\bigg] f_{\scriptsize{\textrm{SR}}}(\omega_i,\omega_s).
\end{align}

We are of course particularly interested in the limit $n \rightarrow \infty$, for which each pump pulse is allowed to traverse the cavity a sufficient number of times so that the intra-cavity pump amplitude becomes extinguished.   In this case, the joint spectral amplitude becomes

\begin{align}
f_{\scriptsize{\textrm{DR}}}(\omega_i,\omega_s) &\equiv
\lim_{n \rightarrow \infty}f_{\scriptsize{\textrm{DR}}}^{(n)}(\omega_i,\omega_s) \nonumber \\
&=t_{1p}\mbox{e}^{i\gamma_p} \bigg[\frac{1+r_{2p}r_{1s}r_{1i}\mbox{e}^{i(\theta_{si}+\theta_p)}}{1-r_{1p}r_{2p}\mbox{e}^{i2\theta_p}}\bigg]  \nonumber \\
& \times f_{\scriptsize{\textrm{SR}}}(\omega_i,\omega_s)
\end{align}

We may then write an expression for the resulting joint spectral intensity for the doubly-resonant nonlinear cavity $S{\scriptsize{\textrm{DR}}}(\omega_i,\omega_s)=|f_{\scriptsize{\textrm{DR}}}(\omega_i,\omega_s)|^2$ as follows

\begin{align}
\label{E:JSIdr}
S_{\scriptsize{\textrm{DR}}}(\omega_i,\omega_s)&= \mathscr{A}_s(\omega_s) \mathscr{A}_i(\omega_i) \mathscr{A}_p(\omega_s+\omega_i)  \nonumber \\
& \times \mathscr{P}(\omega_i,\omega_s) |f(\omega_i,\omega_s)|^2
\end{align}

\noindent where $\mathscr{A}_s(\omega)$ and $\mathscr{A}_i(\omega)$ have already been defined and where $\mathscr{A}_p(\omega)$ is an Airy function associated with the pump mode, and is given as follows

\begin{align}
\mathscr{A}_p(\omega)=\frac{|t_{1p}|^2}{(1-|r_{1p} r_{2p} |)^2}\cdot\frac{1}{1+ \mathscr{F}_p \sin^2(\Delta_p(\omega)/2)}
\end{align}

\noindent in terms of the coefficient of finesse for the pump pulse $\mathscr{F}_p$

\begin{equation}
\label{ }
\mathscr{F}_p=\frac{4|r_{1p} r_{2p}|}{(1-|r_{1p}r_{2p}|)^2}
\end{equation}

\noindent and the phase factor $\Delta_p(\omega)$

\begin{equation}
\label{ }
\Delta_p(\omega)=2\theta_p+\delta_{1p}+\delta_{2p}.
\end{equation}

In Eq.~\ref{E:JSIdr},  $\mathscr{P}(\omega_i,\omega_s)$ represents a phase balancing factor, which is given by

\begin{align}
\label{ }
&\mathscr{P}(\omega_i,\omega_s)=(1+|r_{2p}|)^2 \nonumber \\ &\times\bigg(1-\frac{4|r_{2p}|}{(1+|r_{2p}|)^2}\sin^2\left[\Delta (\omega_i,\omega_s)/2\right] \bigg)
\end{align}

\noindent in terms of the phase term

\begin{equation}
\Delta(\omega_i,\omega_s)=\theta_s+\theta_i+\theta_p+\delta_{1s}+\delta_{1i}+\delta_{2p}.
\end{equation}

From Eq.~\ref{E:JSIdr}, in order to maximize photon pair emission probability at a given frequency pair $\{\omega_{i0},\omega_{s0}\}$, we must ensure that: i) the cavity is resonant for the signal and idler modes; this corresponds to maximizing the Airy functions $\mathscr{A}_s(\omega)$ and $\mathscr{A}_i(\omega)$ at the frequencies $\omega_{s0}$ and $\omega_{i0}$, ii) the cavity is resonant for the pump mode; this corresponds to maximizing the Airy function  $\mathscr{A}_p(\omega)$ at the central pump frequency, and iii) the phase term $\Delta(\omega_{i0},\omega_{s0})$ is made equal to an integer multiple of $\pi$.  Physically, the last condition is fulfilled when the photon-pair amplitude corresponding to a given pass of the pump pulse through the cavity is in phase with the photon-pair amplitude corresponding to the subsequent pass of the pump pulse through the cavity.   Note that the mirror reflection phases $\delta_{\nu \mu}$ (with $\mu=p,s,i$ and $\nu=1,2$) may be chosen so that conditions i) through iii) are fulfilled at given desired frequencies for the three modes.  Of course this translates into the engineering challenge of fabricating mirrors with specific reflection phases for each of the modes, which in principle can be achieved with layered dielectric mirrors where the values for all layer thicknesses are appropriately optimized. \cite{klemens06}.

\begin{figure}[h]
 \includegraphics[width=3.4in]{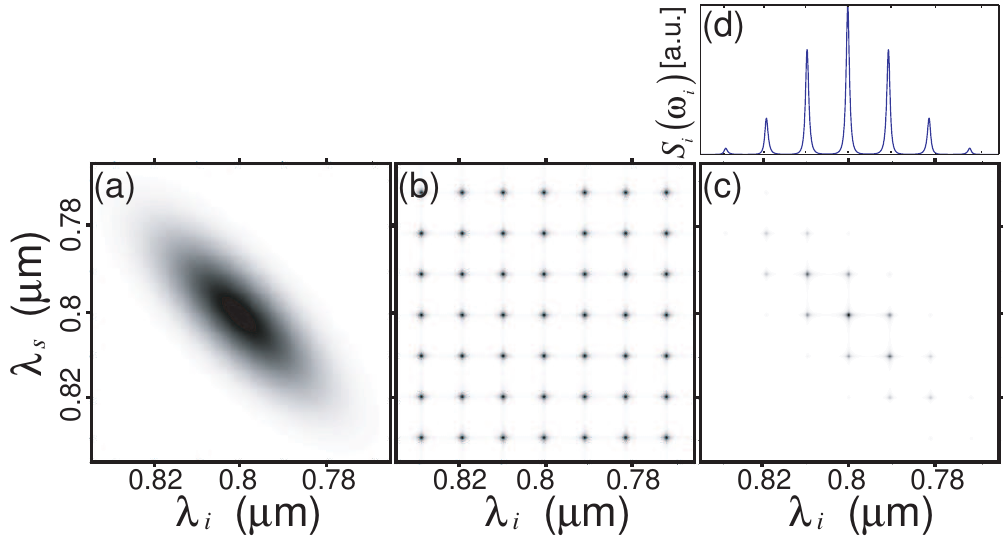}
 \caption{Plotted as a function of the signal and idler frequencies: (a) Joint spectral intensity for the corresponding singly-resonant nonlinear cavity, (b) Pump-mode Airy function $\mathscr{A}_p(\omega_s+\omega_i)$, (c) Phase-balancing function $\mathscr{P}(\omega_i,\omega_s)$, (d) Resulting joint spectral intensity for the doubly-resonant cavity.  (e) Idler-mode, single-photon spectrum. \label{Fg:DRC}}
\label{Fg:DC}
\end{figure}

In order to illustrate the discussion for the doubly-resonant cavity, let us present a specific example.  We will assume a source identical to the example considered in the singly-resonant cavity section (see Fig.~\ref{Fg:JSISC}), except that the pump reflectivity for mirror $2$ is made equal to unity, and that for mirror $1$ is made equal to $|r_{1p}|=0.5$.
While a relatively low value for the pump-mode reflectivity is assumed for this illustration, in a realistic implementation a value much closer to unity is likely to be used. Fig.~\ref{Fg:DC}(A) shows the joint spectral intensity for the equivalent singly-resonant cavity, given by $\mathscr{A}_s(\omega_s)\mathscr{A}_i(\omega_i)|f(\omega_i,\omega_s)|^2$.  Fig.~\ref{Fg:DC}(B) shows the function $\mathscr{A}_p(\omega_s+\omega_i)$, composed of equal-height peaks oriented diagonally (each maximized at a constant value of $\omega_s+\omega_i$).   Fig.~\ref{Fg:DC}(C) shows the function $\mathscr{P}(\omega_i,\omega_s)$, with a sinusoidal structure along the direction defined by $\omega_s+\omega_i$ in $\{\omega_i,\omega_s\}$ space. Fig.~\ref{Fg:DC}(D) shows the resulting joint spectral intensity for the doubly-resonant cavity.  Fig.~\ref{Fg:DC}(E) shows the idler-mode, single-photon spectrum given by the marginal distribution $\int d \omega_s |f_{\scriptsize{\textrm{DR}}}(\omega_i,\omega_s)|^2$.

\begin{figure}[]
 \includegraphics[width=3.4in]{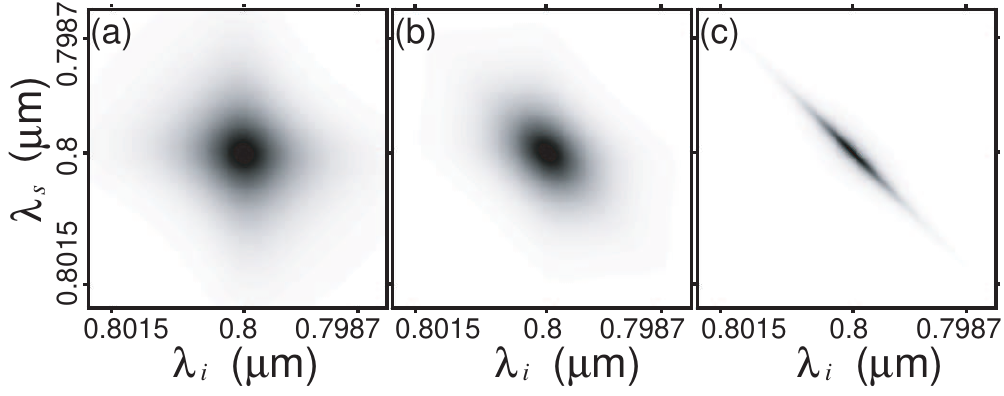}
\caption{Here we show the central cavity-allowed mode as a function of the the signal and idler frequencies for the following values of the pump-mode reflectivity for mirror 1: $|r_{1p}|=0.3$ (panel a), $|r_{1p}|=0.65$ (panel b), and $|r_{1p}|=0.95$ (panel c).  Note that in other respects, this source is identical to that presented in Fig.~\ref{Fg:DRC}.}
\label{Fg:islas}
\end{figure}

Let us note that in the singly-resonant nonlinear cavity case, unless the joint spectral function $f(\omega_i,\omega_s)$ (obtained without a cavity) is unusually narrow-band, the resulting cavity-allowed modes will be fully determined by the product of Airy functions $\mathscr{A}_s(\omega_s)\mathscr{A}_i(\omega_i)$.     This implies that if a single cavity-allowed mode is isolated with an appropriate spectral filter, the resulting two-photon state will be factorable, i.e. it will involve no spectral entanglement.   The situation can be different for the doubly-resonant cavity where functions $\mathscr{A}_p(\omega_s+\omega_i)$ and $\mathscr{P}(\omega_i,\omega_s)$ are oriented diagonally in $\{\omega_i,\omega_s\}$ space.  In particular, if the pump-mode coefficient of finesse $\mathscr{F}_p$ value is considerable, the diagonally-oriented peaks associated with the function $\mathscr{A}_p(\omega_s+\omega_i)$ will imply that the resulting cavity-allowed modes become elongated in such a way that they exhibit spectral anti-correlation.  In the limit of a perfect cavity for the pump (i.e. $\mathscr{F}_p \rightarrow \infty$), strict spectral correlations between the signal and idler modes will be observed.    This behavior is illustrated in Fig.~\ref{Fg:islas}, where the three panels show the central (i.e. centered at the degenerate signal and idler frequency) cavity-allowed spectral mode for the same source as in Fig.~\ref{Fg:DC}, except that the pump coefficient of finesse  is chosen to take the values $2.45$, or $|r_{1p}|=0.3$, (panel A), $21.22$, or $|r_{1p}|=0.65$, (panel B), and $1520$, or $|r_{1p}|=0.95$, (panel C).  It is clear from the figure that the spectral correlations become more pronounced as the pump-mode coefficient of finesse is increased.

\section{Photon-pair rate of emission}

As we have studied above, an important effect of the presence of an optical cavity around the nonlinear medium can be a drastic reduction of the emission bandwidth, at the cost of increasing the correlation time.    In this section we study the effect of the presence of the cavity on the source brightness, defined as the number photon pairs emitted per pump pulse.    We will study this effect both in the singly-resonant and doubly-resonant cases.

It may be shown that the source brightness is given by the following expression

\begin{equation}
B=\frac{b U}{\sigma} \int d \omega_i \int d \omega_s  \left[\frac{k'(\omega_s)\omega_s}{n^2(\omega_s)}\right] \left[\frac{k'(\omega_i)\omega_i}{n^2(\omega_i)}\right] S(\omega_i,\omega_s)
\label{E:brightness}
\end{equation}

\noindent where $U$ is the pump pulse energy, the function $S(\omega_i,\omega_s)$ is the joint spectral intensity to be replaced by the function $S_{\scriptsize{\textrm{SR}}}(\omega_i,\omega_s)$ for the singly-resonant cavity case and by the function $S_{\scriptsize{\textrm{DR}}}(\omega_i,\omega_s)$ for the doubly-resonant cavity case; $b$ is a function of experimental parameters such as the nonlinear cavity length and the crystal nonlinearity; a prime denotes a frequency derivative.

Let us first discuss the singly-resonant nonlinear cavity case.    From section~\ref{Sec:SRC}, we know that the presence of the cavity leads to the factor $\mathscr{A}_s(\omega_s)\mathscr{A}_i(\omega_i)$ in the joint spectral amplitude.    This factor means that photon-pair emission takes place at well-defined, cavity-allowed spectral modes.  While the width of the peaks which define the cavity-allowed modes decreases with the coefficient of finesse (see Eq.~\ref{Eq:deltaomega}), the height of the peaks, given by $|t_{2\mu}|^2/(1-|r_{2 \mu}|)^2=(1+|r_{2 \mu}|)/(1-|r_{2 \mu}|)$, increases with the reflectivity of mirror $2$ (or equivalently with the coefficient of finesse).  This means that if the pump envelope function $\alpha(\omega_i+\omega_s)$ overlaps one or more of the cavity-allowed modes in $\{\omega_i,\omega_s\}$ space, the resulting probability amplitude for frequencies within one of these modes can be much higher than it would be without the cavity.   However, for a fixed pump pulse energy $U$, as the pump bandwidth is increased, the pump energy must be distributed over a greater range of frequencies; if some of these pump frequencies $\omega_p$ with an associated locus $\omega_p=\omega_s+\omega_i$  in $\{\omega_i,\omega_s\}$ space do not overlap cavity-allowed modes, they do not contribute to photon pair emission.  A consequence of this is that for a sufficiently large pump bandwidth ($\sigma \gtrsim \Delta \omega /\sqrt{ 2 \ln2 }$, where $\Delta \omega$ is the spectral separation between cavity-allowed modes, see Eq.~\ref{Eq:Deltaw}), the lower probability amplitude obtained without a cavity over a large bandwidth results in essentially the same source brightness compared with the higher probability amplitude obtained within narrowband cavity-allowed modes, obtained in the presence of the cavity.

Importantly, this means that, for a sufficiently large pump bandwidth, the joint spectral intensity given by function $|f(\omega_i,\omega_s)|^2$ (obtained without a cavity) is re-distributed in $\{\omega_i,\omega_s\}$ space by the presence of the cavity to yield the cavity-modified joint spectral intensity $S_{\scriptsize{\textrm{SR}}}(\omega_i,\omega_s)$, \textit{without} a resulting reduction in the emitted flux.  This should be compared with an identical source, except with the cavity removed, where photon pairs are transmitted through a filter with a spectral transmittance function which matches that associated with the nonlinear cavity previously considered.  Mathematically, this corresponds to a transmittance function  $\mathscr{A}(\omega_s)\mathscr{A}(\omega_i)$, normalized to a unit height.  Note that this could be implemented with an appropriate (empty) optical cavity placed following the nonlinear crystal.   In this case, only those frequency pairs which match the allowed modes are retained and other frequency pairs are discarded.   This of course leads to a considerable reduction of the photon-pair flux.  Note that in the limit where the filter-allowed modes are monochromatic (i.e. for a high-finesse cavity filter), the transmitted flux approaches zero, because a vanishing portion of the two photon state matches the filter-allowed modes.  Thus, the essential advantage of incorporating the cavity at the source itself, is that rather than generating a certain bandwidth and retaining only specific narrow-band modes (the situation which corresponds to standard filtered SPDC), we can engineer the source so that only certain narrow-band modes are generated in the first place. This leads to an important flux enhancement.

However, the use of a singly-resonant nonlinear cavity permits us not only to maintain, but in fact it permits us to exceed, the flux obtained with an equivalent source without a cavity (while of course reducing the SPDC emission bandwidth).    Indeed, if as the coefficient of finesse is increased leading to a reduction of the bandwidth of the cavity-allowed modes, the pump bandwidth is also reduced so that $\sigma \approx \delta \omega/\sqrt{ 2 \ln2}$ is fulfilled, then all the frequencies present in the pump pulses contribute to photon pair generation at the higher probability amplitudes possible within the cavity-allowed modes.    This implies an important source brightness enhancement, over the brightness associated with an equivalent source without a cavity.  Furthermore, this enhancement is controlled by the reflectivity of mirror $2$ : higher values of the reflectivity lead to a greater brightness enhancement.  As we will see below, if the pump bandwidth $\sigma$ is selected appropriately, a singly-resonant cavity can lead to a dramatic source brightness enhancement in the high-finesse limit where $|r_2| \rightarrow 1$.

Let us consider for reference the source brightness dependence vs pump bandwidth for the equivalent source without a cavity. If all the frequencies present in the pump pulses are phasematched (i.e. if the phasematching function is broader than the pump envelope function), then as the pump bandwidth is increased, function $|f(\omega_i,\omega_s)|$ becomes broader (along the direction $\omega_s+\omega_i$ in $\{\omega_i,\omega_s\}$ space) linearly with $\sigma$. The factor $1/\sigma$ in Eq.~\ref{E:brightness} then implies that within the phasematching bandwidth (for $\sigma \lesssim \sigma_{pm}$ where $\sigma_{pm}$ is the phasematching bandwidth), the source brightness shows essentially no dependence on the pump bandwidth, or equivalently on the pump pulse duration.  Note that this insensitivity to pump bandwidth means that for SPDC the number of photon pairs produced depends on the incident pump energy, rather than on the instantaneous pump power.

\begin{figure}[]
 \includegraphics[width=3.4in]{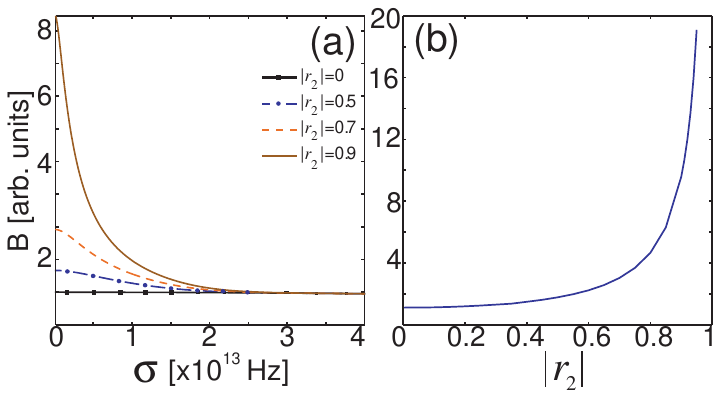}
\caption{(a)  This plot shows the source brightness, as a function of the pump bandwidth, for different values of the SPDC mode mirror $2$ reflectivity $|r_2|$. Note that $|r_2|=0$ corresponds to the equivalent source without a cavity. (b)  This plot shows the optimum source brightness (for $\sigma \rightarrow 0$) as a function of mirror $2$ reflectivity $|r_2|$.}
 \label{Fg:SRC}
\end{figure}

In order to illustrate the above discussion, let us consider the source of figure~\ref{Fg:JSISC}, except that the pump bandwidth and the mirror $2$ reflectivity are allowed to vary.    Figure~\ref{Fg:SRC}(A) shows the source brightness vs pump bandwidth, calculated numerically,  for the equivalent source without cavity (black curve with square markers) showing, as expected, essentially no dependence on $\sigma$.   Note that the numerical brightness calculations presented here assume that the factors $k' \omega/ n^2$ in Eq.~\ref{E:brightness} are well approximated by the corresponding numerical values when evaluated at the central SPDC emission frequency.  The curves are also normalized so that unity corresponds to the source brightness for the equivalent source without cavity, for small $\sigma$.  Figure~\ref{Fg:SRC}(A) also shows the dependence of the source brightness on the pump bandwidth, calculated numerically,  for a singly-resonant cavity with three different values for the mirror 2 reflectivity ($|r_2|=|r_{2s}|=|r_{2i}|=0.5,0.7,0.9$; the corresponding curves from low to high values are colored blue (dot-dashed line), orange (dashed line) and brown (continuous line) respectively).  It is evident from the figure that as the pump bandwidth is decreased in value, at a certain point, the source brightness for the singly-resonant cavity begins to exceed the brightness of the equivalent source without a cavity.  At this point the pump bandwidth corresponds to the spectral separation between cavity modes, i.e $\sigma=\Delta \omega/\sqrt{ 2 \ln2}$.  As the pump bandwidth $\sigma$ is reduced further, around the value for $\sigma$ which corresponds to the width of a cavity-allowed mode, the source brightness reaches a plateau.  The maximum attainable brightness (i.e. within this plateau) is controlled by the reflectivity of mirror $2$ (or equivalently by the coefficient of finesse of the cavity): this enhancement increases as the mirror $2$ reflectivity is increased. The resulting dependence of the optimum source brightness vs mirror $2$ reflectivity is shown in Figure~\ref{Fg:SRC}(B). For $|r_2|$ approaching unity, this enhancement scales as $1/(1-|r_2|)$, so that in principle the brightness can increase without limit upon increasing the singly-resonant cavity coefficient finesse and reducing the pump bandwidth correspondingly.  Of course, in practice this will be limited by how large the coefficient of finesse (or the mirror $2$ reflectivity) can be made.

The above discussion makes it clear that a singly-resonant nonlinear cavity can lead to an important source brightness enhancement (coupled with a drastic reduction in the emission bandwidth) compared to the equivalent source without a cavity.   It is to be expected that if the cavity is doubly-resonant so that each pump pulse traverses the cavity multiple times contributing to the photon pair emission probability with each pass, the source brightness should increase further.  In what follows, we will explore the behavior of the source brightness with respect to experimental parameters of the doubly-resonant nonlinear cavity.

The joint spectral amplitude for a doubly-resonant nonlinear cavity differs from that for the singly-resonant nonlinear cavity in the appearance of the factor $ \mathscr{A}_p(\omega_s+\omega_i) \mathscr{P}(\omega_i,\omega_s)$.  The multiple passes of the pump pulses through the cavity are reflected in the magnitude of the numerical factor in front of the Airy function $\mathscr{A}_p(\omega_s+\omega_i)$, which increases with the coefficient of finesse for the pump.  As we have already discussed in section~\ref{Sec:CDR}, increasing the pump coefficient of finesse also leads to the appearance of signal-idler spectral correlations, within a given cavity-allowed mode, as governed by the width of the Airy function $\mathscr{A}_p(\omega_s+\omega_i)$.  While each cavity-allowed mode has a factorable spectral distribution in the singly-resonant cavity case, it may have a highly elongated shape for the doubly-resonant case, where the degree of elongation is controlled by the pump coefficient of finesse.    The reduced width of the cavity-allowed modes along the direction $\omega_s+\omega_i$ in $\{\omega_i,\omega_s\}$ space implies that the pump bandwidth used for the optimum SPDC brightness should be reduced as the pump coefficient of finesse is increased (otherwise, some pump frequency components will not contribute to photon-pair generation, reducing the attainable brightness).     We note that this is consistent with the fact that for a sufficiently large pump bandwidth (or correspondingly, for sufficiently short pump pulses as compared to the cavity round-trip time), and for a high finesse cavity, the incident pump pulses will tend to be reflected off mirror $1$ of the nonlinear cavity before photon-pair emission can take place, thus essentially suppressing photon-pair generation.

In order to illustrate the effect of making the cavity resonant for the pump mode, in addition to the signal and idler modes, figure~\ref{Fg:DRC}(A) shows the source brightness vs the reflectivity of mirror $1$ for the pump mode, for three different choices of the pump bandwidth $\sigma$ ($\sigma=2\times 10^{11},3.5\times 10^{11},5\times 10^{11}$Hz).  Here, a source characterized by $r_{1p}=0$ approaches the singly-resonant case (however, it is not exactly equivalent since the doubly-resonant cavity with $r_{1p}=0$ leads to two passes through the crystal instead of only one for the singly-resonant cavity).  The curves have been normalized to the source brightness obtained at $r_{1p}=0$.    The source brightness enhancement due to the double resonance is clear from Fig.~\ref{Fg:DRC}. The value $\sigma=5\times 10^{11}$Hz assumed for one of the curves corresponds roughly to the largest pump bandwidth within the area which yields the optimum bandwidth for the corresponding singly-resonant cavity [see Figure~\ref{Fg:SRC}(A)].  As we expect from the discussion in the previous paragraph, a reduction of the pump bandwidth matching the reduction of the cavity-allowed mode width along the direction $\omega_s+\omega_i$ in $\{\omega_i,\omega_s\}$ space leads to an increased source brightness.  Of course, in the limit where the cavity is perfectly resonant for the pump mode, i.e. where $r_{1p}=1$, then the pump pulse cannot enter the cavity in the first place, and the brightness must drop to zero.  This expected drop in brightness is shown in figure~\ref{Fg:DRC}(B), which is similar to panel A, except that it concentrates on very large reflectivities $|r_{1p}|$.

\begin{figure}[]
 \includegraphics[width=3.4in]{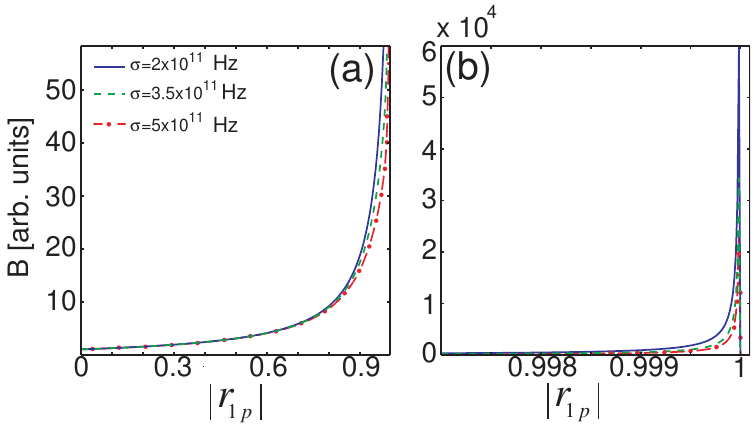}
 \caption{(a) Source brightness as a function of the mirror $1$ reflectivity for the pump $|r_{1p}|$, for three different values of the pump bandwidth ($\sigma=2\times 10^{11},3.5\times 10^{11}$ and $5\times 10^{11}$Hz).  . (b) Similar to panel (a), for large values of $|r_{1p}|$; note the expected drop in source brightness near $|r_{1p}|=1$. Note that the vertical plotting range in panels a and b differs by a factor of 1000.}
 \label{Fg:DRC}
\end{figure}

\section{A specific source design}
\label{Sec:SpDesign}

In this section we present a `recipe' for the design of a photon-pair source based on a singly-resonant cavity, where one of the two SPDC modes can be coupled to a specific transition of an atomic system.  Note that this implies matching both the central frequency and bandwidth of the atomic transition to those of those of one of the two SPDC modes.  We also provide a specific example as an illustration for the $D_{5/2}-P_{3/2}$ transition of the ${}^{40}\mbox{Ca}^{+}$ ion, which is centered at $854.2$nm, with a width of $\sim 20$MHz.   The recipe consists of the following six steps.
\begin{enumerate}

\item The frequency for the signal photon is selected to match the central frequency of the desired atomic transition. In our illustration, this central frequency is $\omega_{s0}=2 \pi c/\lambda_s$, with $\lambda_s=854.2$nm.

\item The central frequency of the pump laser is selected according to experimental constraints.   In our illustration, we assume a central pump frequency $\omega_{p0}=2 \pi c/\lambda_{p0}$, with $\lambda_{p0}=400$nm.  This also defines the central idler frequency, $\omega_{i0}=\omega_{p0}-\omega_{s0}$, corresponding to a wavelength of $\lambda_{i}=752.26$nm.

\item Once the central frequencies for all three modes are known, we select a specific type of nonlinear crystal and select the crystal cut angle so that phasematching is fulfilled at these frequencies, in other words so that $k_p(\omega_{p0})-k_s(\omega_{s0})-k_i(\omega_{i0})=0$.  In our illustration we use a BBO crystal with type-I phasematching, resulting in a crystal cut angle $\theta_{pm}=29.1^\circ$.

\item  We select the nonlinear cavity length and therefore also the crystal length (with $L=\ell$) through Eq.~\ref{Eq:Deltaw} so that the spectral separation between two subsequent cavity modes is greater than a certain threshold $\Delta \lambda_{max}$ which is considered to represent the smallest spectral separation that permits the isolation through appropriate spectral filtering of a single cavity-allowed mode.  In our illustration we assume $\Delta \lambda_{max}=0.5$nm, which leads to a cavity length of $L=220 \mu$m .

\item Equating the bandwidth of the atomic transition (in our case $2 \pi \times 20$MHz in angular frequency) to the signal-mode SPDC bandwidth defines through Eq.~\ref{Eq:deltaomega} the required mirror $2$ reflectivity.  In our illustration, this leads to a mirror $2$ reflectivity $|r_2|=0.9999$, corresponding to a coefficient of finesse given by $\mathscr{F}=4\times10^{8}$.  Note that while this is undoubtedly a large finesse, it is within current experimental capabilities.

\item In order to take advantage of the brightness enhancement made possible by the singly-resonant cavity, we select the pump bandwidth $\sigma$ so that $\sigma \lesssim \delta \omega/\sqrt{2 \ln 2}$, which defines the spectral width of each cavity-allowed mode.   In our illustration, this corresponds to $\sigma \lesssim 106.7$ MHz.   Note that if the pump bandwidth is greater than this value, the spectral characteristics of the emitted photon pairs will not be affected, though the source brightness will be lower than the largest attainable value.

\end{enumerate}

\begin{figure}[]
 \includegraphics[width=3.4in]{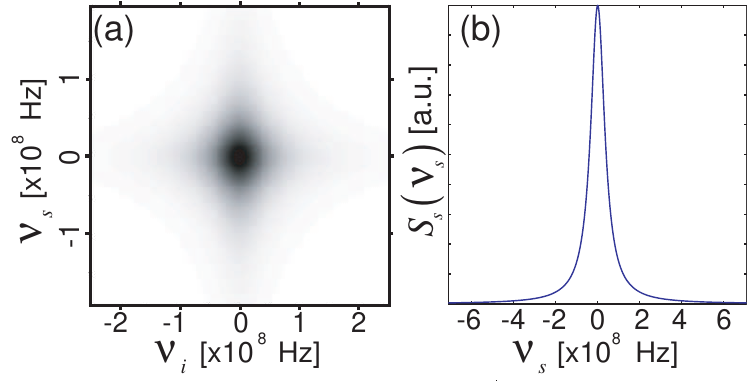}
\caption{(a) Joint spectral intensity associated to the cavity-allowed mode of interest  (b) Signal-mode, single-photon spectrum.  These plots show the signal photon centered at $854.2$nm, with a bandwidth of $20$MHz.
 \label{Fg:SpecExample}}
\end{figure}

Figure~\ref{Fg:SpecExample}(A) shows the joint spectral intensity, plotted as a function of the spectral detunings $\nu_s=\omega_s-\omega_{s0}$ and $\nu_i=\omega_i-\omega_{i0}$, for the above illustration.  Figure~\ref{Fg:SpecExample}(B) shows a plot of the corresponding signal-mode single-photon spectrum.  As required, the signal photon is centered at $\lambda=854.2$nm, and the idler photon is centered at $\lambda=743.9$nm.  Also, the bandwidth of the signal mode is, as required, close to $20$MHz.  Although of course the source brightness could be increased further by making the cavity doubly-resonant, we have not included that possibility in this illustration.

\section{Conclusions}

We have studied the generation of photon pairs through the process of spontaneous parametric downconversion in a nonlinear cavity, formed by a nonlinear crystal surrounded by mirrors.  We have studied two different regimes: i) that of a singly-resonant cavity where the signal and idler modes are resonant in the cavity, and ii) that of a doubly-resonant cavity where the pump is also resonant in the cavity.   In both cases we have studied the spectral properties of the generated photon pairs as a function of the nonlinear cavity parameters.    We have shown that in the presence of the cavity, emission takes place in spectrally well-defined modes.  These modes are characterized by a width which scales as $\mathscr{F}^{-1/2}$ where $\mathscr{F}$ is the coefficient of finesse for the SPDC photons.  Thus, a high-finesse cavity leads to narrowband photon-pair emission, which may be exploited for the implementation of photon-pair sources tailored for coupling with a specific transition of an atomic system.   We have shown that this is achieved at the cost of an increased photon-pair correlation time:  in the limit of a perfect cavity, the emission modes lose their photon-pair character since the signal and idler photons are no longer correlated in time.

We have also explored the effect of the cavity on the source brightness: we have shown that for a sufficiently large pump bandwidth, the source brightness remains unchanged with respect to the equivalent source without a cavity.  However, we have shown that if the pump bandwidth is small compared to the spectral separation between cavity-allowed modes, the source brightness is enhanced with respect to the equivalent source without a cavity, and  this enhancement is a monotonically-increasing function of the coefficient of finesse.    We have shown that making the cavity doubly-resonant leads to the appearance of spectral correlations in each of the cavity-allowed modes; in the limit of a perfect cavity (for the pump), the signal and idler photons become perfectly correlated in frequency.  We have shown that a doubly-resonant cavity can result in an enhanced source brightness with respect to the equivalent singly-resonant cavity; this is due to the fact that the pump pulse can now traverse the nonlinear medium multiple times, in each pass contributing to the photon-pair emission probability.   As the coefficient of finesse for the pump mode is increased, the pump bandwidth must be reduced accordingly in order to reach the optimum brightness.    Considering: i) the additional engineering challenge of building a doubly-resonant, vs a singly-resonant cavity, and ii) the very substantial source brightness enhancement attainable with a singly-resonant cavity, experimental implementations could reasonably concentrate on the singly-resonant case.  We believe that this paper will be useful for the design of narrowband photon-pair sources, a key component for efficient light-matter interfaces.

\section*{Acknowledgements}

This work was supported in part by: (i) CONACYT, Mexico; (ii) DGAPA, UNAM; (iii) FONCICYT project no. 9142, and (iv) Intelligence Advanced Research Projects Activity (IARPA) under Army Research Office ARO Contract No. W911NF-05-1-0397.



\begin{thebibliography}{26}


\bibitem{mosley08} P.J. Mosley, J.S. Lundeen, B.J. Smith, P. Wasylczyk, A.B. U'Ren, C.Silberhorn and I.A. Walmsley, Phys. Rev. Lett. \textbf{100}, 133601 (2008)

\bibitem{odonnell07} K.A. O'Donnell and A.B. U'Ren, Optics Letters \textbf{32} 817 (2007)

\bibitem{herzog94} T. J. Herzog, J. G. Rarity, H. Weinfurter, and A. Zeilinger, Phys. Rev. Lett. \textbf{72}, 629  (1994)

\bibitem{konig05} F. Konig, E. J. Mason, F. N. C. Wong, and M. A. Albota,
Phys. Rev. A \textbf{71}, 033805 (2005)

\bibitem{fedrizzi07} A. Fedrizzi, T. Herbst, A. Poppe, T. Jennewein, and A.
Zeilinger, Opt. Express \textbf{15}, 15377 (2007)

\bibitem{kuklewicz06}C. E. Kuklewicz, F. N. C. Wong, and J. H. Shapiro,
Phys. Rev. Lett. \textbf{97}, 223601 (2006)

\bibitem{wang04} H. Wang, T. Horikiri, and T. Kobayashi,  Phys. Rev. A \textbf{70}, 043804 (2004).

\bibitem{neergard-nielsen07} J. S. Neergaard-Nielsen, B. Melholt Nielsen, H. Takahashi,
A. I. Vistnes and E. S. Polzik, Opt. Exp. \textbf{15}, 7940 (2007)

\bibitem{haase08} A. Haase, N. Piro, J. Eschner and M.W. Mitchell, Opt. Lett. \textbf{34}, 55 (2009)

\bibitem{ou99} Z. Y. Ou and Y. J. Lu, Phys. Rev. Lett. \textbf{83}, 2556 (1999)

\bibitem{lu00} Y. J. Lu and Z. Y. Ou, Phys. Rev. A \textbf{62}, 033804 (2000)

\bibitem{andrews01} R. Andrews, E.R. Pike and S. Sarkar, Appl. Opt. \textbf{40}, 4050 (2001)

\bibitem{hariharan00} P. Hariharan and B.C. Sanders, J. Mod. Opt. \textbf{47} 1739 (2000)

\bibitem{klemens06} G. Klemens and Y. Fainman, Opt. Exp. \textbf{14} 9864 (2006)




\end{thebibliography}
\end{document}